\documentclass[sigconf, nonacm=true]{acmart}

\usepackage{multirow}
\usepackage{algpseudocode}
\usepackage{algorithm}
\usepackage{graphicx}
\usepackage{subcaption}
\usepackage{array}
\usepackage[export]{adjustbox}
\usepackage{flushend}

\makeatletter
\g@addto@macro{\UrlBreaks}{\UrlOrds}
\makeatother

\usepackage{graphicx}
\graphicspath{ {./figures/} }

\usepackage[htt]{hyphenat}

\begin{document}

\title{What Did It Look Like: A service for creating website timelapses using the Memento framework}


\author{Dhruv Patel}
\affiliation{
  \department{Department of Computer Science}
  \institution{Old Dominion University}
  \city{Norfolk}
  \state{VA}
  \postcode{23529}
  \country{USA}
}
\email{dpate006@odu.edu}

\author{Alexander C. Nwala}
\affiliation{%
  \department{Observatory on Social Media}
  \institution{Indiana University}
  \city{Bloomington}
  \state{IN}
  \country{USA}
}
\email{anwala@iu.edu}

\author{Michael L. Nelson}
\orcid{0000-0003-3749-8116}
\affiliation{%
  \department{Department of Computer Science}
  \institution{Old Dominion University}
  \city{Norfolk}
  \state{VA}
  \postcode{23529}
  \country{USA}
}
\email{mln@cs.odu.edu}

\author{Michele C. Weigle}
\orcid{0000-0002-2787-7166}
\affiliation{%
  \department{Department of Computer Science}
  \institution{Old Dominion University}
  \city{Norfolk}
  \state{VA}
  \postcode{23529}
  \country{USA}
}
\email{mweigle@cs.odu.edu}

\begin{abstract}
 Popular web pages are archived frequently, which makes it difficult to visualize the progression of the site through the years at web archives. The What Did It Look Like (WDILL) Twitter bot shows web page transitions by creating a timelapse of a given website using one archived copy from each calendar year. Originally implemented in 2015, we recently added new features to WDILL, such as date range requests, diversified memento selection, updated visualizations, and sharing visualizations to Instagram. This would allow scholars and the general public to explore the temporal nature of web archives.
\end{abstract}

\maketitle

\section{Introduction}

The World Wide Web is impermanent and its rapid growth along with constant changes has made preservation of past information valuable. The Web has become an integral part of our lives and has doubled in size every year since 2012 \cite{internet-size}. Because of this, web archives are essential resources for research in fields such as humanities and social science \cite{arora-15, milligan16}. There are various web archiving organizations such as the Internet Archive\footnote{\url{https://archive.org/index.php}}, UK Web Archive\footnote{\url{https://www.webarchive.org.uk/ukwa/}}, Perma.cc\footnote{\url{https://perma.cc/}}, and many more that have captured billions of past web pages. 

\begin{figure}[ht]
\centering
\includegraphics[width=0.45\textwidth]{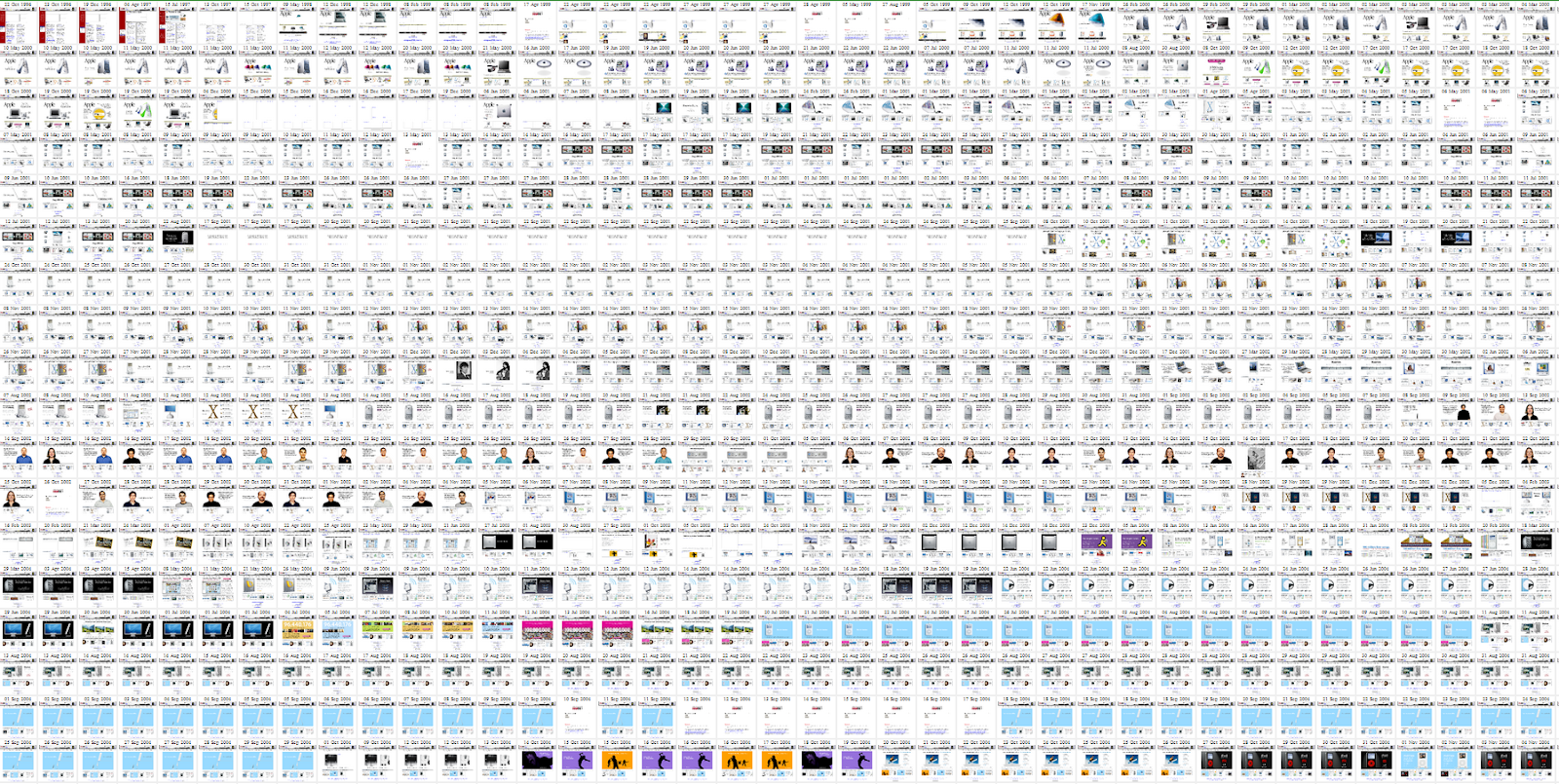}
\caption{First 700 Mementos for \url{https://www.apple.com} spanning 1997 to 2006 \cite{alsum-ecir14}}
\label{fig:700-apple-thumbnails}
\end{figure}

Popular web pages are archived frequently, some even multiple times a day. These archived web pages would have to be manually explored one at a time to view web page transitions overtime and this could be a very tedious process. Figure \ref{fig:700-apple-thumbnails} shows archived web pages of \url{https://www.apple.com} from 1997 to 2006. The grid view makes it slightly easy to visualize the web page changes however it does not show the time frame of each archived web page. This makes it difficult to know the time frame of the transitions. Including all archived versions in the grid view would make it more difficult to view all web page transitions.

As the goal is to see a web page change over a long period of time, a common way to view such changes is via a timelapse. With people already frequently sharing images and videos on social media, this makes social media platforms an easy way to share timelapses to a large audience. In 2015, the What Did It Look Like service was introduced to create and share generated timelapses of archived web pages to Twitter and Tumblr. The service is implemented as a Twitter Bot that accepts timelapse requests through tweets and responds to users. The goal of the project was to add enhancements such as date range requests, diversified memento selection, updated visualizations, and sharing the timelapses to Instagram.

\section{Viewing Mementos of a Webpage Over Time}

A snapshot of a saved web page at a given time is known as a \textit{memento} \cite{memento-framework}. Mementos consist of various resources from the live web page such as text, media, and CSS. A \textit{TimeMap} is a collection of mementos for a specific URL. This enables for reconstruction of past or inactive web pages that can be used to view transitions of web pages.

Visualizing mementos with significant changes achieves the goal of capturing the transitions of a web page. In order to determine mementos with significant changes, the manual process involves comparing the current memento with the previous one in the TimeMap. AlSum found that the most effective way to measure similarities between web pages is by comparing computed SimHashes \cite{simhash} on the HTML content \cite{alsum-ecir14}. This technique has been utilized by a service called TimeMap Visualization (TMVis) to create visualizations that capture web page changes over time \cite{tmvis-arxiv-2020}. TMVis calculates a Hamming Distance between the computed SimHashes of a pair of mementos. A user specified threshold is used to determine if the memento is considered to have changed significantly. This is done by evaluating if the Hamming Distance is greater than the threshold. A higher threshold will lead to fewer but significantly different mementos. For a lower threshold, the opposite is true. This method requires downloading each memento's HTML content and this could be a time consuming process for large TimeMaps. A similar functionality is provided as part of the GLAM-Workbench \cite{glam-workbench}.

\begin{figure}[ht]
\centering
\includegraphics[width=0.45\textwidth]{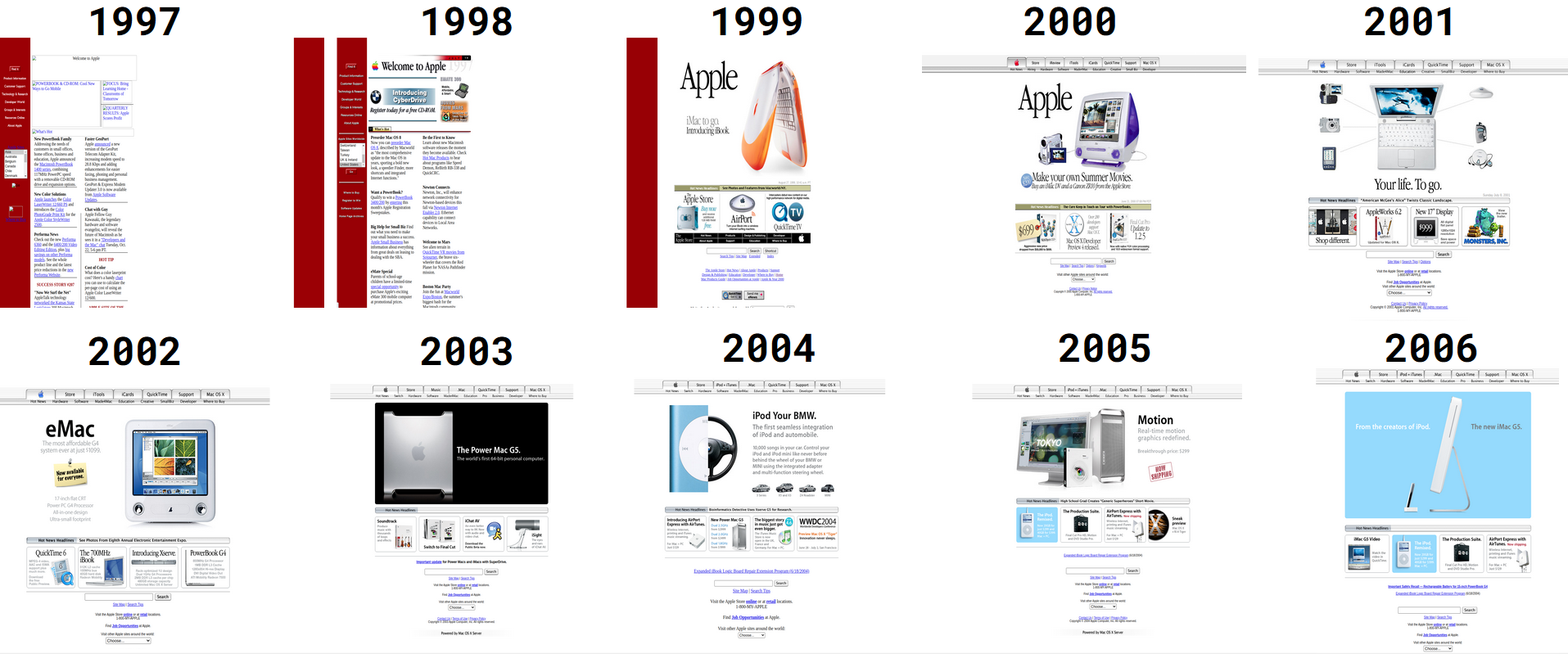}
\caption{One Memento per year for \url{https://www.apple.com} spanning 1997 to 2006 \cite{alsum-ecir14}}
\label{fig:memento-per-year}
\end{figure}

An alternate way to choose mementos for the timelapse is by picking one memento per year. This is a faster option where mementos are selected based on the time and not their content. Figure \ref{fig:memento-per-year} demonstrates this as it shows that the web page design from 1997 to 1999 are quite similar and a change occurred later in 2000.

\section{Sharing visualizations using Twitter Bots}
Twitter is one of the platforms that permits bots given that they comply with a set of rules. Despite certain limitations from the Twitter platform, there is still a wide range of functionality offered for automated accounts. A study from the University of Southern California and Indiana University in 2017 estimated that approximately 9\% to 15\% of accounts on Twitter are bots \cite{varol2017online}. A Twitter Bot is a great way for a program to interact with a live audience and it is also easily accessible given the size of the user base on the platform. Some examples of popular Twitter Bots include @EmojiAquarium\footnote{\url{https://twitter.com/EmojiAquarium}} which posts a randomly generated aquarium using emojis every few hours, @year\_progress\footnote{\url{https://twitter.com/year_progress}} that generates a progress bar for the current year, and @netflix\_bot\footnote{\url{https://twitter.com/netflix_bot}} which tweets daily about new releases on Netflix Instant. To easily allow users to submit timelapse requests for websites, What Did It Look Like is implemented as a Twitter Bot. The service also notifies the user when the request is received and when the timelapse is generated.

\section{Design and Implementation}

What Did It Look Like (WDILL) was originally developed by Alexander Nwala in 2015 \cite{wdill-blog}. This section details the overall design and implementation of the original service followed by enhancements added in the new version.

\subsection{Original WDILL Service}

The What Did It Look Like (WDILL) service relies on users sending requests of websites to create timelapses of by tweeting \texttt{\#whatdiditlooklike URL}. WDILL is a time-based service and executes at the specified times. Once the service executes, it accesses the Twitter API to fetch requests and processes them.. Next, TimeMaps for the processed requests are acquired using an external TimeMap Aggregator. These TimeMaps are used to select mementos to create visualizations. The generated visualizations are then posted to Tumblr\footnote{\url{https://tumblr.com/}}. Figure \ref{fig:process_overview} below shows the process overview of WDILL.

\begin{figure*}[ht]
\centering
\includegraphics[width=0.65\textwidth]{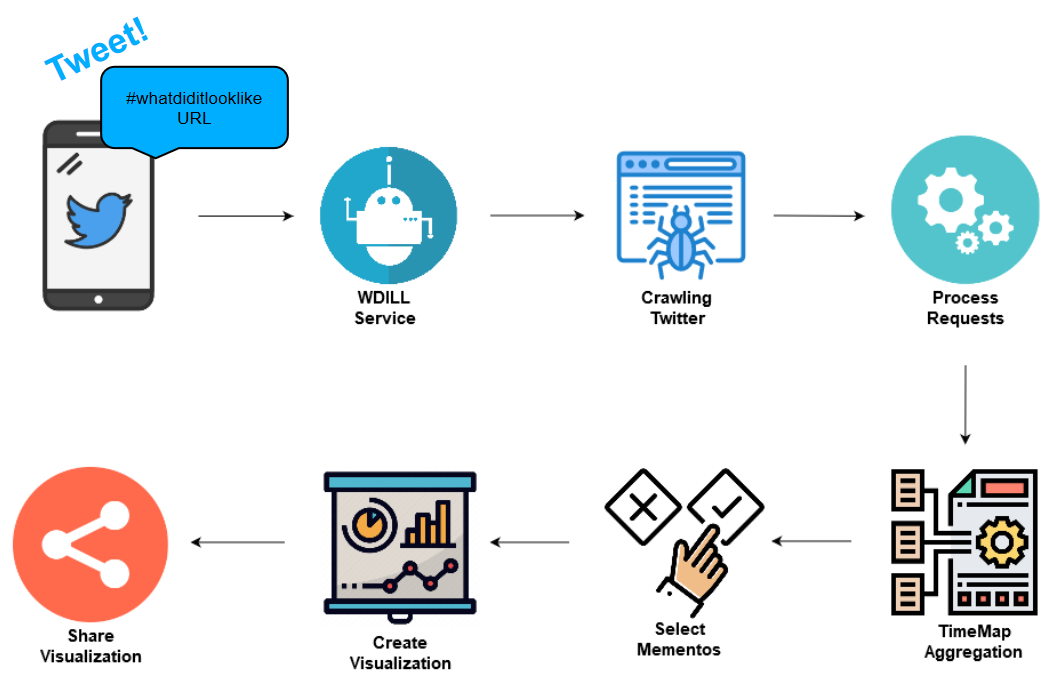}
\caption{What Did It Look Like Process Overview}
\label{fig:process_overview}
\end{figure*}

\subsection{Creating Requests}

WDILL uses Twitter, a social media platform with a large user base to allow users to easily submit timelapse requests. In order to request for a timelapse of a website, users can simply tweet \texttt{\#whatdiditlooklike URL} and they will be notified when the request has been received by WDILL. WDILL allows users to submit multiple URLs in the same tweet. This can be done by writing URLs in a comma delimited list. Figure \ref{fig:request_tweet_multiURL} below shows an example of a timelapse request.

\begin{figure}[ht]
\centering
\includegraphics[width=0.45\textwidth]{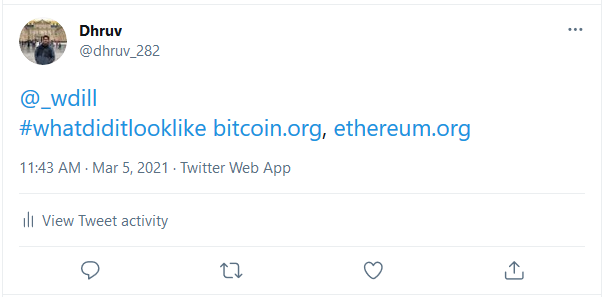}
\caption{A request for a multiple URL timelapse}{ \url{https://twitter.com/dhruv_282/status/1361807543983624196}}
\label{fig:request_tweet_multiURL}
\end{figure}

\subsection{Processing Requests}

In order to obtain requests, WDILL uses Twitter's API to search for tweets that contain \texttt{\#whatdiditlooklike} \cite{twitter-api}. The ID of the latest request during the previous search is stored in a file, \texttt{timelapseTwitterSinceIDFile.txt}. This ID is used as a parameter of the Twitter API call to obtain tweets after that ID to avoid creating duplicate requests. If the file is empty, all tweets with the hashtag are obtained and processed.

All links in the tweet are expected to be positioned right after \texttt{\#whatdiditlooklike}. The links are wrapped in Twitter's t.co format \cite{twitter-tco}. These t.co links redirect to URLs shared by the user and these URLs could contain further redirects. To account for this, the parsed t.co links are expanded to acquire the actual URL requested by the user.

\begin{figure}[ht]
\centering
\includegraphics[width=0.45\textwidth]{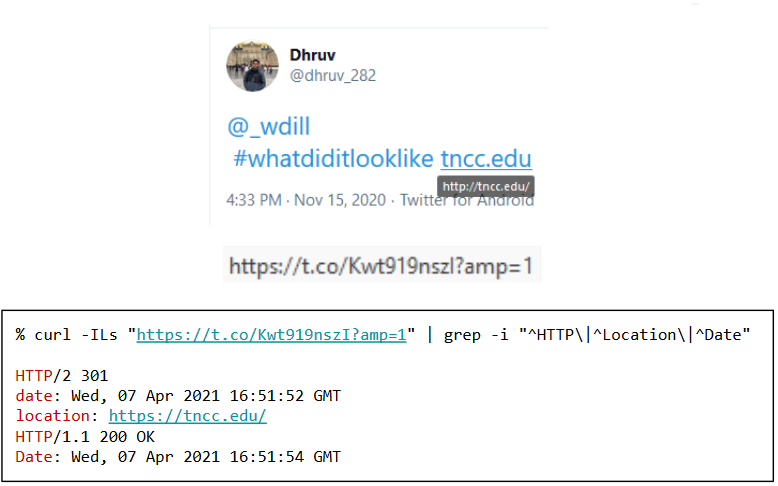}
\caption{An example of t.co link redirection}{\url{https://twitter.com/dhruv_282/status/1328088640623423494}}
\label{fig:tco-redirect-example}
\end{figure}

As it is possible for requested URLs to have been processed and fulfilled in the past, the new requests are processed only if they are requested after a certain number of days to avoid creating similar content. Here the date of the last fulfilled request with the same URL is obtained. The number of days between the last fulfilled request and the new request is evaluated. This difference of days is compared with a parameter in the \texttt{config} file, \texttt{nominationDifferential} to determine if the content in the fulfilled request is still fresh. If the number of days is greater than the \texttt{nominationDifferential}, the content in the last fulfilled request is considered to be outdated and the user is notified that the URL has been queued for processing. Otherwise, the content from the fulfilled request is considered to be fresh and the user is notified with the amount of days after the URL can be resubmitted and a link to the timelapse of the last fulfilled request. The default value of the \texttt{nominationDifferential} parameter is 30 days, which means that the user can resubmit the same URL after 30 days.

\begin{figure*}[ht]
\centering
\includegraphics[width=0.75\textwidth]{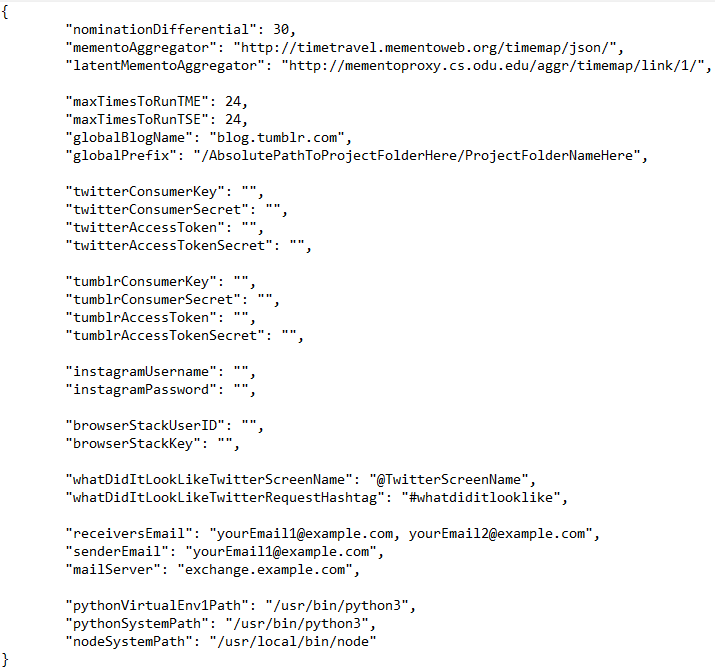}
\caption[]{Sample of config file}
\label{fig:sample-config}
\end{figure*}

\begin{figure}[ht]
\centering
\includegraphics[width=0.45\textwidth]{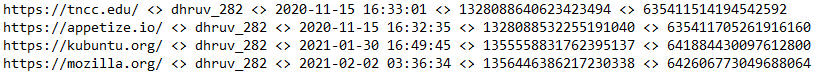}
\caption[]{Sample of nominated requests that have been processed}
\label{fig:processed_requests}
\end{figure}

Selected URLs are formed into requests and are stored in a file, \texttt{twitter\_requests\_wdill.txt}. A request consists of a URL, Twitter user ID, tweet datetime, and tweet ID. These values are delimited by \texttt{`<>'} in the file. A sample of the file is illustrated in Figure \ref{fig:processed_requests}

\subsection{TimeMap Aggregation}

Mementos of websites can be found in varying amounts and quality across different web archives. TimeMaps for each request in \texttt{twitter\_requests\_wdill.txt} are acquired using an external TimeMap Aggregation service. The URL specified in the \texttt{mementoAggregator} field in the \texttt{config} file is the TimeMap Aggregator used. By default, WDILL uses Time Travel's ``Do It Yourself'' TimeMap aggregation service to acquire TimeMaps from different web archives \cite{timetravel-apis}.

\begin{figure}[ht]
\centering
\includegraphics[width=0.45\textwidth]{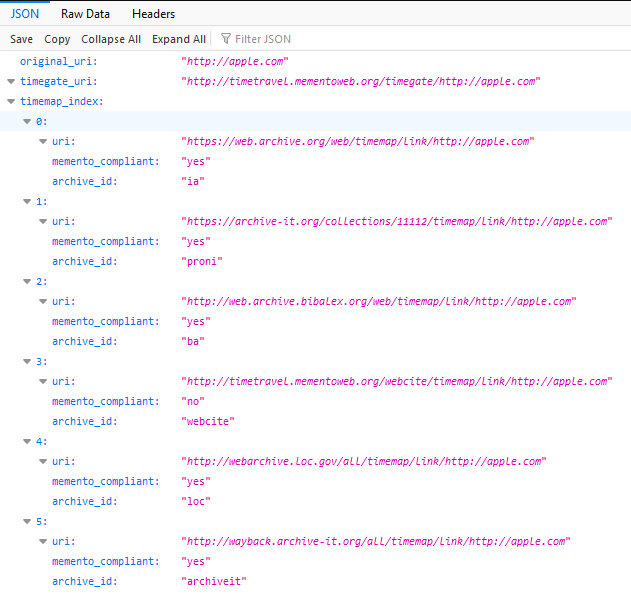}
\caption{TimeMaps for \url{http://apple.com} from Time Travel's TimeMap Aggregator}{\url{http://timetravel.mementoweb.org/timemap/json/http://apple.com}}
\label{fig:timemap_aggregator_sample}
\end{figure}

As shown in Figure \ref{fig:timemap_aggregator_sample}, not all web archives are compliant with the Memento Framework \cite{memento-framework}. WDILL acquires mementos of all TimeMaps returned from the aggregator regardless of the value in the \texttt{memento\_compliant} field. The Memento framework's link format \cite{memento-framework} is expected while parsing TimeMaps. Mementos from each TimeMap are stored in separate lists and each item in the list consists of a memento URL and the datetime delimited by ``\texttt{*+*+*}''. These TimeMap lists containing memento information are stored in one parent list.

\subsection{Memento Selection}

Once all TimeMaps are parsed, mementos are selected to create a timelapse. In order to capture the transition of a web page over time, one memento from each year is selected. This is done by using an associative array data structure which holds key value pairs and does not allow duplicate key values. The key in this context is the year and value is the memento link (URI-M). For each memento, its year is checked for in the associative array. If that year returns a URI-M, then the current memento is ignored otherwise, the URI-M of the current memento is added.

\subsection{Creating Visualizations}

In order to create visualizations, screenshots of the selected mementos are taken. PhantomJS is used to load each memento in a headless browser to take screenshots \cite{phantomjs}. Screenshots are saved in PNG format and resized to a 1024x768 resolution. The naming convention for these files is the year of the memento, i.e. \texttt{YYYY.png}. A folder named with the MD5 hash of the requested URL is created to store all media related to the request. Once a screenshot of a memento has been successfully captured, the memento's year and link is stored in a file named \texttt{urlsFiles.txt} in the folder with the screenshots. 

ImageMagick's \texttt{convert} function is used to add labels at the top of the image on screenshots indicating the year of the memento. The images are used to create a GIF representing a timelapse of the requested URL with a 4 second animation interval between each frame \cite{convert}. Gifsicle is used to optimize the size of the GIF using a high improved compression ratio \cite{gifsicle}.

\subsection{Sharing Visualizations}

The generated GIFs are shared to a Tumblr blog using PyTumblr API \cite{pytumblr}. Values for API credentials and blog page specified in the \texttt{config} file.

\begin{figure}[ht]
\centering
\includegraphics[width=0.45\textwidth]{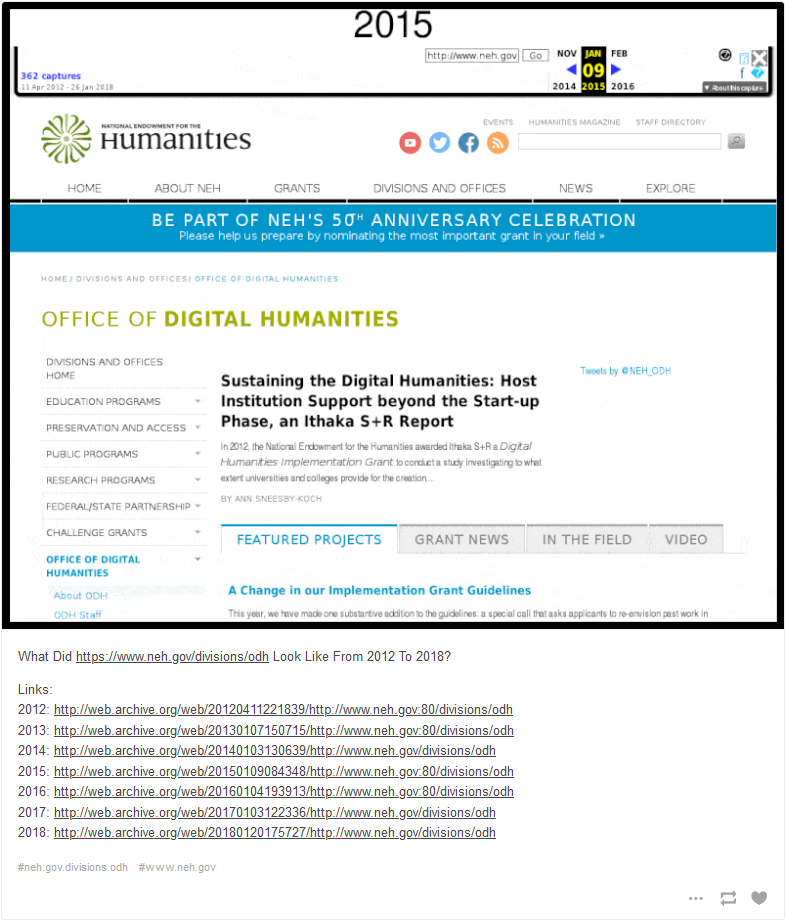}
\caption{Tumblr post for a timelapse of \url{https://www.neh.gov/divisions/odh}}{\url{https://whatdiditlooklike.mementoweb.org/tagged/neh.gov.divisions.odh}}
\label{fig:old_tumblr_post}
\end{figure}

After the post has been published on the blog, a message containing a link to the timelapse post is issued as a reply to the original request tweet. A separate status update is also made on Twitter account's timeline to notify the account followers about a new timelapse.

\begin{figure}[ht]
\centering
\includegraphics[width=0.45\textwidth]{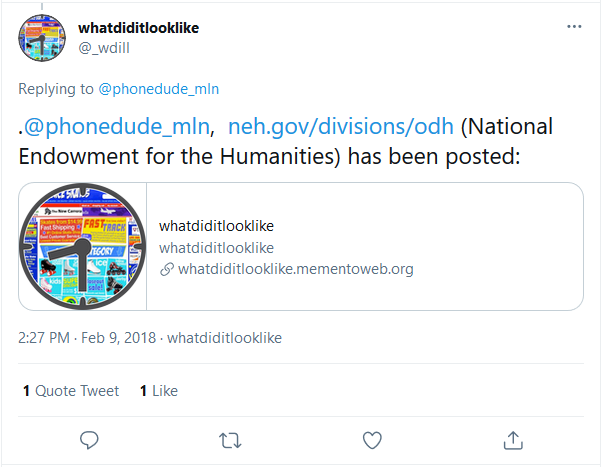}
\caption{Response to the request tweet}{\url{https://twitter.com/_wdill/status/962045423983640579}}
\label{fig:old_reply_tweet}
\end{figure}

As a clean up procedure, all fulfilled requests are moved to a file named \texttt{twitter\_requests\_wdill\_store.txt} and the directories containing media related to requests are renamed to the request URL's hash concatenated with the datetime the request was fulfilled.

\begin{figure}[ht]
\centering
\includegraphics[width=0.45\textwidth]{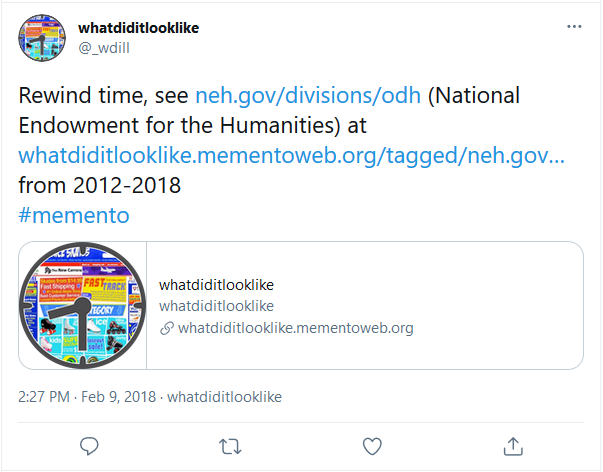}
\caption{Post on timeline for followers}{\url{https://twitter.com/_wdill/status/962045403091759104}}
\label{fig:old_post_tweet}
\end{figure}

\subsection{Original Implementation}

Originally, the program was implemented in Python 2 and utilized PhantomJS to capture memento screenshots \cite{wdill-blog}. The service was executed at four random times a day generated by a script written in Python. This original implementation is located in a GitHub repository at \texttt{\url{https://github.com/oduwsdl/wdill/tree/cc97ad6a919b6ca8c2803a404636ecb00c10495a}}. 

The service begins by executing the \texttt{timelapseTwitter.py} program which fetches new tweets containing \texttt{\#whatdiditlooklike}. The tweet ID of the last fetched tweet in the previous run is saved in a file \texttt{timelapseTwitterSinceIDFile.txt} in order to keep track of the newer tweets. This program obtains the text after the hashtag expecting it to be a URL or a comma delimited list of URLs. Extracted URLs will be wrapped in Twitter's t.co format so this program will expand the t.co links to obtain the actual URL tweeted by the user. The user is notified if the requested URL has been nominated and if not, then links to the latest output are included in the message. URLs may not be nominated if timelapses for those URLs have been posted within the last 30 days. Nominated URLs and relevant information such as username, tweet ID, and timestamp are written to a file named \texttt{twitter\_requests\_wdill.txt}.

\begin{figure*}[ht]
\centering
\includegraphics[width=0.75\textwidth]{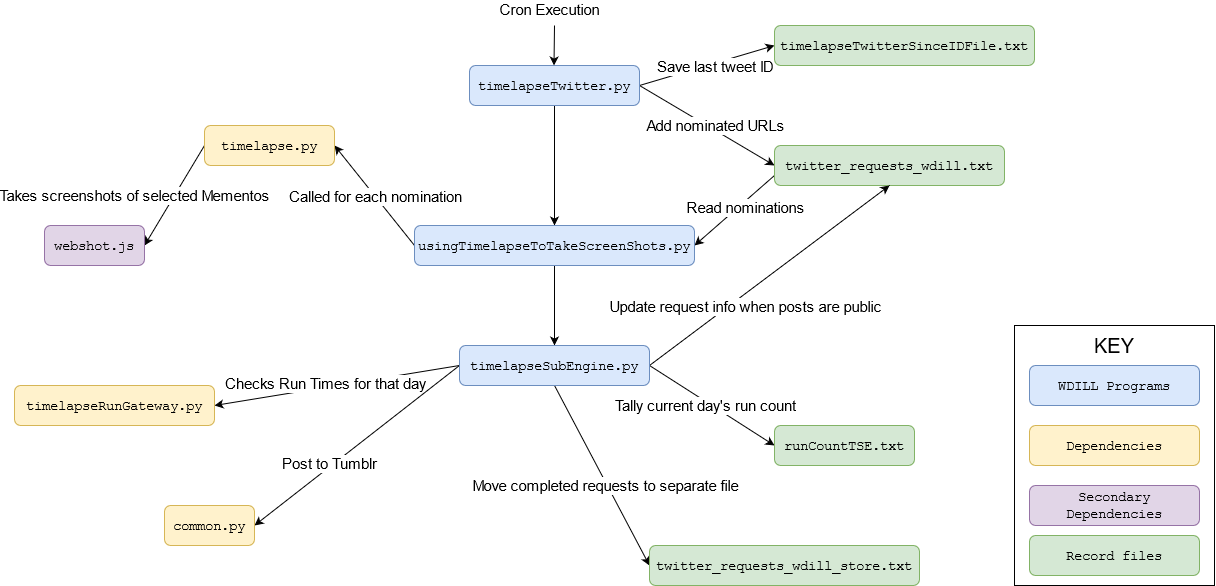}
\caption{Original WDILL program structure}
\label{fig:old-program-execution-structure}
\end{figure*}

The \texttt{usingTimelapseToTakeScreenShots.py} is called next and it supplies each request in \texttt{twitter\_requests\_wdill.txt} to \texttt{timelapse.py}. \texttt{timelapse.py} creates a folder to store files related to the nomination, fetches TimeMaps, and selects mementos. The URLs of selected mementos are stored in a file \texttt{urlsFile.txt} located in a folder created for the nomination. A hash of the canonicalized URL is used as a name for the folder. The process calls \texttt{webshot.js} to capture screenshots for all URLs written to the \texttt{urlsFile.txt}. The year of the mementos are added to the appropriate screenshot and they are used to create GIF timelapses.

After processing all nominations, \texttt{timelapseSubEngine.py} is called to post the timelapses to a specified Tumblr blog. The currently deployed demo service posts to a blog hosted on the MementoWeb site\footnote{\url{https://whatdiditlooklike.mementoweb.org}}. This program also keeps track of the run count of executions for the each day. A limit to run counts is specified in the \texttt{config} file in order to control the amount of content published. The \texttt{runCountTSE.txt} file stores the daily run counts of the service. Once the content is published, the user is notified about this and a separate tweet is posted on the timeline. The current service utilizes the @\_wdill\footnote{\url{https://twitter.com/_wdill}} Twitter account for interaction with the service and users.

\subsection{Enhancements}

WDILL has achieved fulfilling the goal of automating timelapse creation upon user requests. The enhancements we have added to the existing service include date range requests, diversified memento selection, updated visualizations, and sharing visualization to Instagram.

\subsection{Date Range Requests}

\begin{figure}[ht]
\centering
\includegraphics[width=0.45\textwidth]{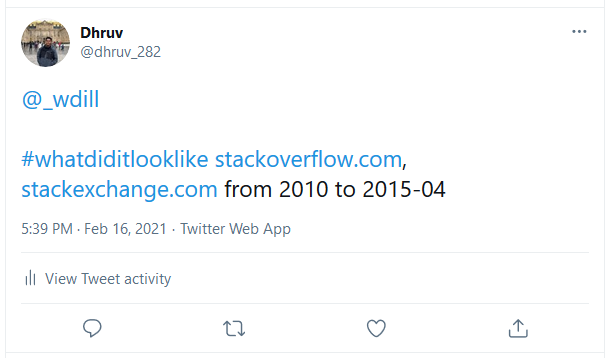}
\caption{A request for multiple URLs within a date range}{\url{https://twitter.com/dhruv_282/status/1361807543983624196}}
\label{fig:request_tweet_daterange_multiURL}
\end{figure}

Users can request for timelapses of websites between a certain duration. WDILL recognizes dates in the ISO 8601 date format \cite{iso-8601}. Supported date formats include \texttt{YYYY}, \texttt{YYYY-MM}, and \texttt{YYYY-MM-DD}. The service allows mixing these formats. For example the user can specify a duration of \texttt{YYYY-MM} to \texttt{YYYY}. If a date range is specified along with multiple URLs, it is applied to all URLs in the tweet. Figure \ref{fig:request_tweet_daterange_multiURL} is an example of a timelapse request for multiple URLs requests within a date range. While processing requests, if a date range is not found or the range is invalid, a date range of \texttt{0 - 0} is set indicating that a date range will not be applied. 

\subsection{Diversified Memento Selection}

After acquiring all TimeMaps, mementos are filtered if a date range is provided in the request. The date range in the request is expected to have the years specified. If the month and date portion are not specified, the values will be filled with the first month and first day of that year. Only mementos that were captured on or between these dates are preserved and the rest are discarded.

\begin{figure}[ht]
\centering
\includegraphics[width=0.35\textwidth]{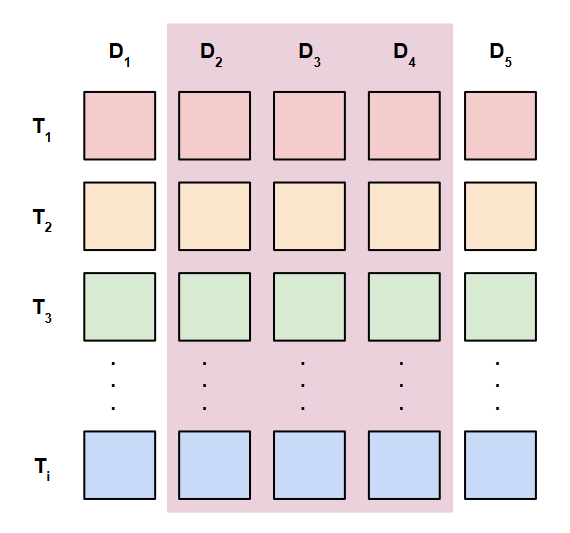}
\caption{Filtering out mementos outside of date range D\textsubscript{1} to D\textsubscript{4}}
\label{fig:daterange_filter}
\end{figure}

\begin{figure}[ht]
\centering
\includegraphics[width=0.4\textwidth]{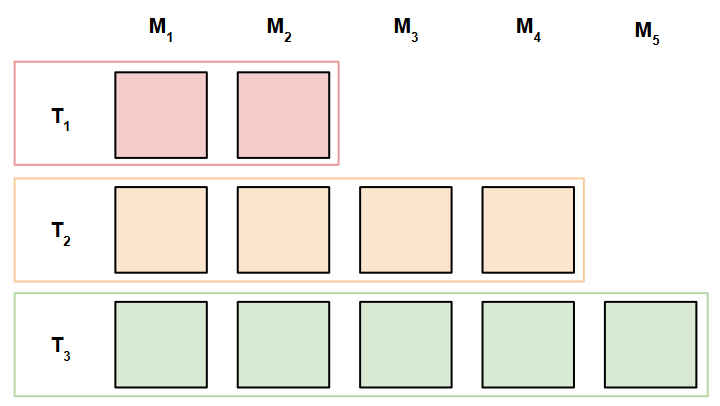}
\caption{Sorting TimeMaps by memento count}
\label{fig:timemap-sort}
\end{figure}

Similar to the original algorithm, only one memento from each year is selected. One of the objectives of this process is to include mementos from various archives. In order to achieve this, the first step is to sort the list of TimeMaps in ascending order by memento count as illustrated in Figure \ref{fig:timemap-sort}. This will allow the first few selections for certain years to be made from smaller TimeMaps. The years with no memento selected will be taken care of by the bigger TimeMaps. 

As the given URL may have been requested in the past, the memento selection algorithm attempts to generate unique visualizations by selecting a different set of mementos. The $N^{th}$ memento from each year is selected where $N-1$ is the number of times the URL has been requested in the past. In order to determine how many times the URL has been requested in the past, \texttt{twitter\_requests\_wdill.txt} and \texttt{twitter\_requests\_wdill\_store.txt} files are searched through. These files contain ongoing and past request information. If there is only one memento for a certain year across all TimeMaps, it will be selected for every request of that URL in order to avoid blank spots in the timeline. Before selecting a potential memento, its status code is checked to verify the page's validity. The next memento is selected if the current potential memento returns an HTTP 404 response.

\begin{figure*}[ht]
\centering
\includegraphics[width=0.75\textwidth]{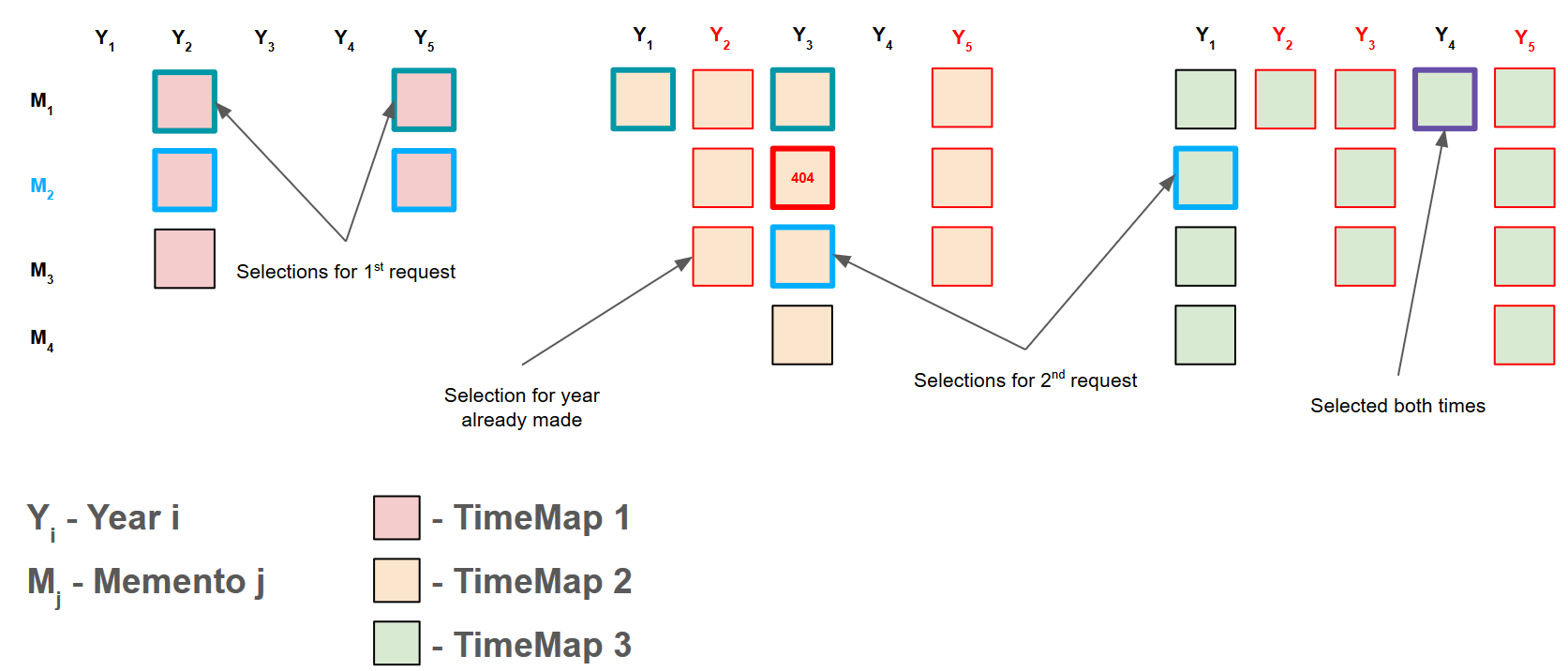}
\caption{New memento selection process}
\label{fig:memento_selection}
\end{figure*}

\subsection{Updated Visualization}

After a screenshot is captured, two transparent watermark labels are applied to the bottom left of the image. The first label indicates the datetime of the memento and the other shows the domain name of the archive. 

\begin{figure}[ht]
\centering
\includegraphics[width=0.45\textwidth]{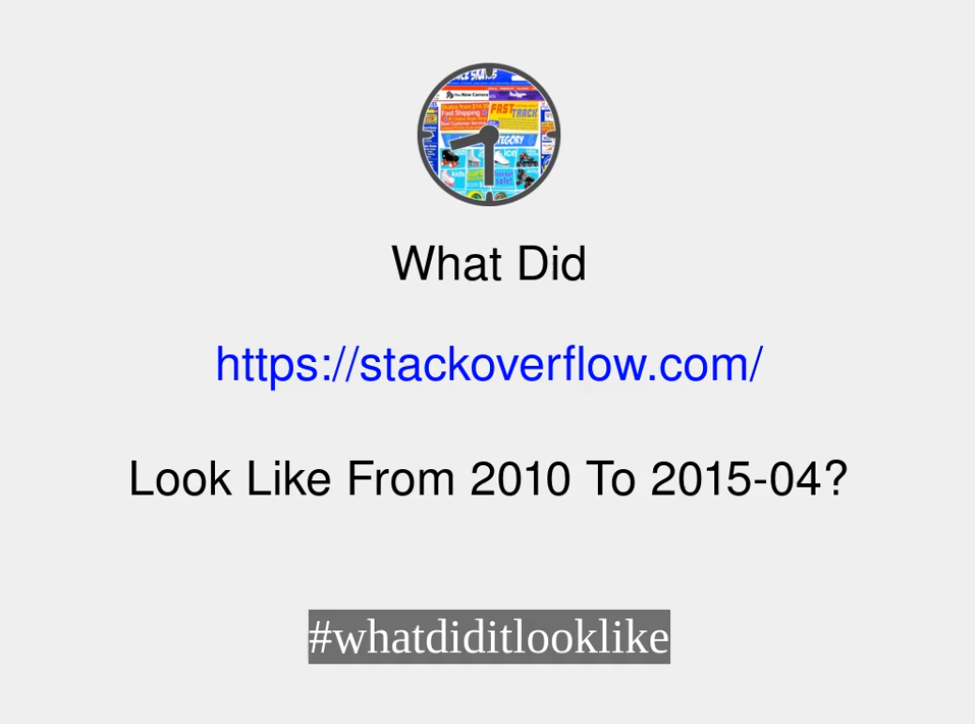}
\caption{Title Slide for a timelapse request with a date range}
\label{fig:title_slide_example}
\end{figure}

Each visualization begins with a title slide that consists of the requested URL and date range of the timelapse as shown on Figure \ref{fig:title_slide_example}. This image is generated before taking screenshots by using the \texttt{wdill\_titleSlide\_generator.sh} script. An MP4 video is generated with the title slide and memento screenshots using FFmpeg \cite{ffmpeg}. MP4 videos require more storage space than GIFs but they provide support for audio. WDILL utilizes royalty free soundtracks from Bensound and Mixkit for the generated MP4 files \cite{bensound, mixkit}.

\begin{figure}[ht]
\centering
\includegraphics[width=0.45\textwidth]{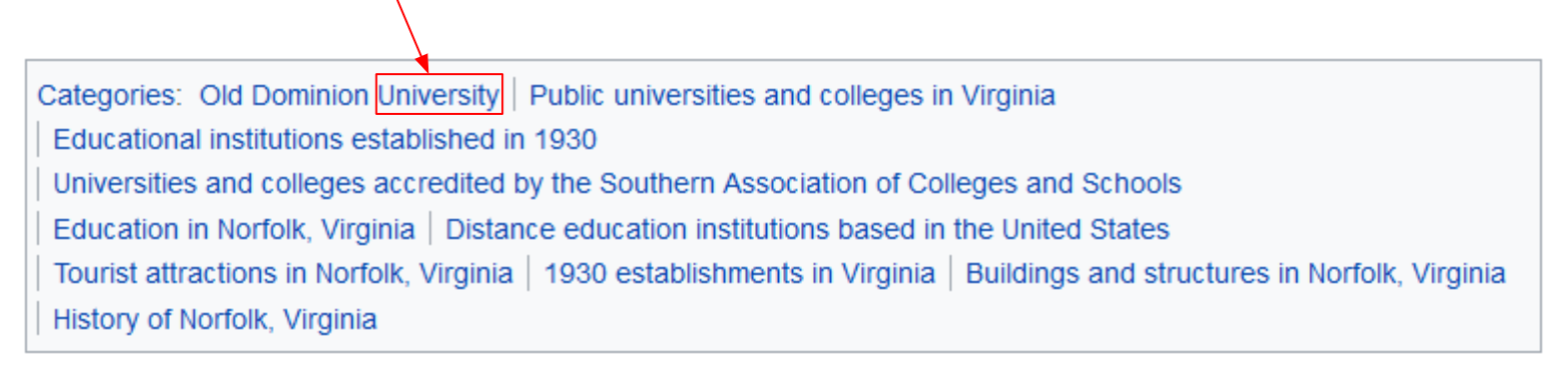}
\caption{Categories of Old Dominion University's Wikipedia page}
\label{fig:wikipedia_category_example}
\end{figure}

WDILL categorizes URLs in order to select a soundtrack from a relevant genre. Pages on Wikipedia are categorized and WDILL accesses this information using Python's Wikipedia API \cite{wikipedia-api}. A search query with the value as the URL is issued to Wikipedia. The results are ordered by query relevance, where the first result most relevant. Categories of the first item in the search results are obtained to determine a generalization of this page. Each generalization has a set of keywords that are searched for in the categories obtained. For example in Figure \ref{fig:wikipedia_category_example}, the word University is recognized in the first category and this page is categorized under education. The URL is categorized as \textit{other} if there are no search results returned or no keyword matches.

\begin{figure}[ht]
\centering
\includegraphics[width=0.45\textwidth]{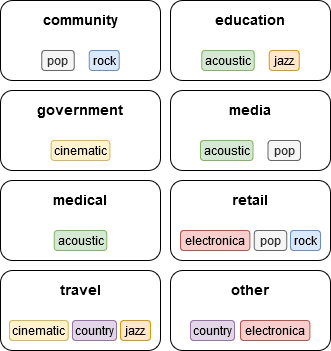}
\caption{Diagram showing genres relevant to categories}
\label{fig:category_genre}
\end{figure}

A random soundtrack from the relevant genre of the URL category is selected. If the category has more than one relevant genre, then a genre is selected at random. A random start point in the sound track is selected and it is verified that the duration from the selected start point to the end is long enough to cover the whole timelapse. This is done to make the visualizations as unique as possible. The mappings from category to genres used by WDILL are shown in Figure \ref{fig:category_genre}.

\subsection{Sharing to Social Media}

Since Tumblr provides MP4 video support, WDILL posts the updated visualization instead of a GIF. The generated MP4 video is also posted to the specified Instagram account in the \texttt{config} file. 

\begin{figure}[ht]
\centering
\includegraphics[width=0.45\textwidth]{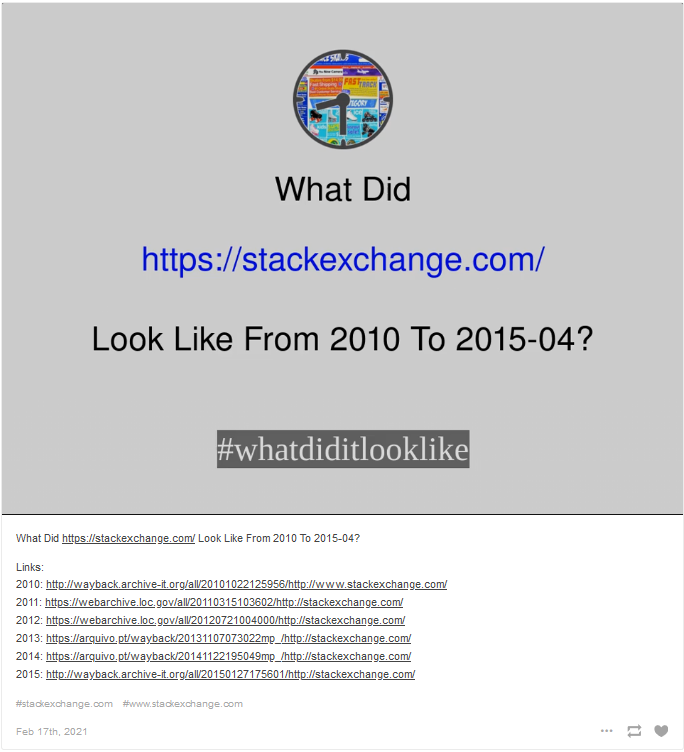}
\caption{A Tumblr post for a timelapse video of \url{https://stackexchange.com}}{\url{https://whatdiditlooklike.mementoweb.org/tagged/stackexchange.com}}
\label{fig:tumblr_post}
\end{figure}

It is not possible to post a GIF on Instagram as it will be converted to an image where only the first frame is uploaded. The solution to this is to post a video timelapse with more than one frame and at least three seconds long \cite{insta-blog}. In order to tackle the challenge of automating Instagram post creation, a script \texttt{instagram.js} was created. This script loads the mobile version of Instagram's website by changing the \texttt{user-agent} on a headless browser and automates navigation using Puppeteer \cite{puppeteer}. 

We later discovered that the Instagram website does not allow for video uploads, which meant that this could only be done using the mobile application. In order to solve this issue, Appium, a mobile navigation framework is used to automate the process on a smartphone \cite{appium}. Information such as an element's ID, class name or xpath is needed to locate and simulate triggers. This information is obtained using UI Automator Viewer, a tool included in Android Studio that collects the current XML dump, takes a screenshot, and maps the XML content to the screenshot.

\begin{figure}[ht]
\centering
\includegraphics[width=0.45\textwidth]{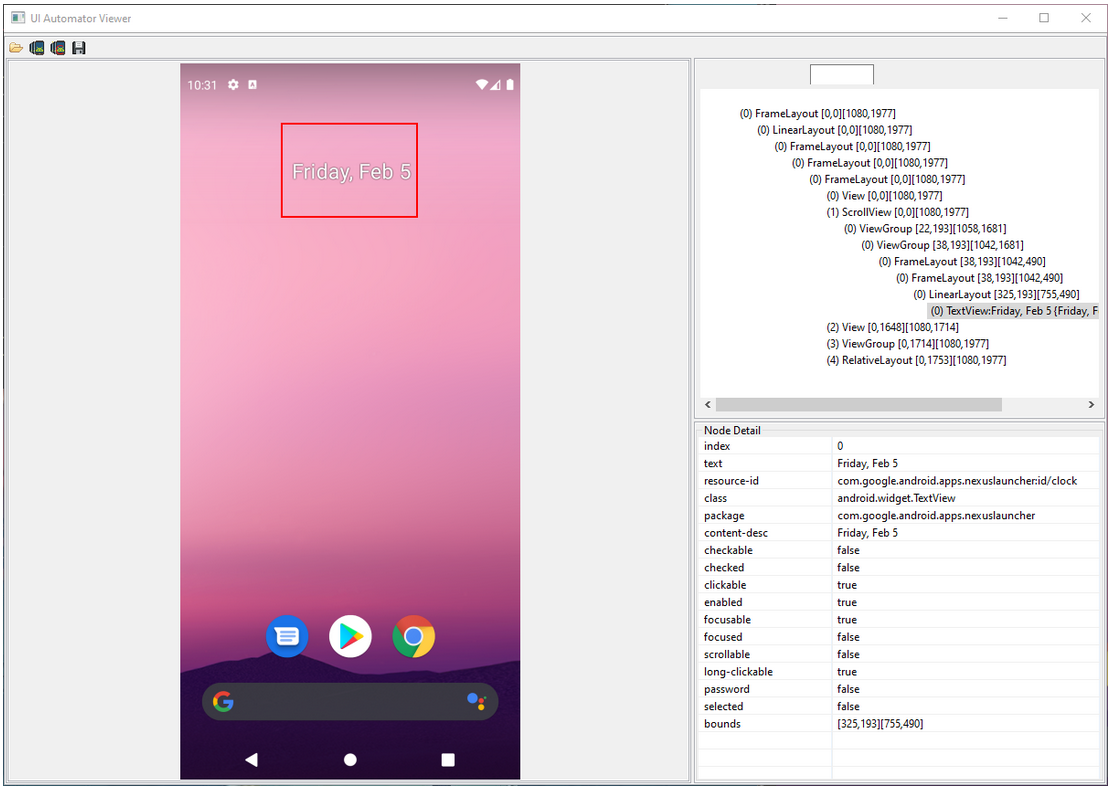}
\caption{Screenshot showing the functionality of UI Automator Viewer}
\label{fig:uiautomatorviewer_sample}
\end{figure}

This tool was used to obtain all necessary information to create an Appium script to automate navigation. In order to avoid having a dedicated Android hardware device to run the Appium script on, a platform called BrowserStack is utilized to get access to a cloud hosted Android device \cite{browserstack}. This platform support the Appium framework and interacts with the program using a REST API. Instagram version 169.3.3.0.30.135's installer file (\texttt{.apk}) is saved since the Appium script was built on this version and any updates to the UI elements may possibly cause errors. The script \texttt{instagramWithBrowserStack.py} uploads the visualization and Instagram APK on BrowserStack, initializes a session, posts to Instagram using Appium, and cleans up by deleting the uploaded files prior to session termination. The script also acquires the link to the post by copying it to the clipboard and fetching clipboard contents from the host. 

The response notification post made to the user includes the GIF timelapse of the requested URL as well as links of the Tumblr and Instagram posts. The GIF timelapse is also included on the post created for the Twitter account timeline to notify the followers.

\begin{figure}[ht]
\centering
\includegraphics[width=0.45\textwidth]{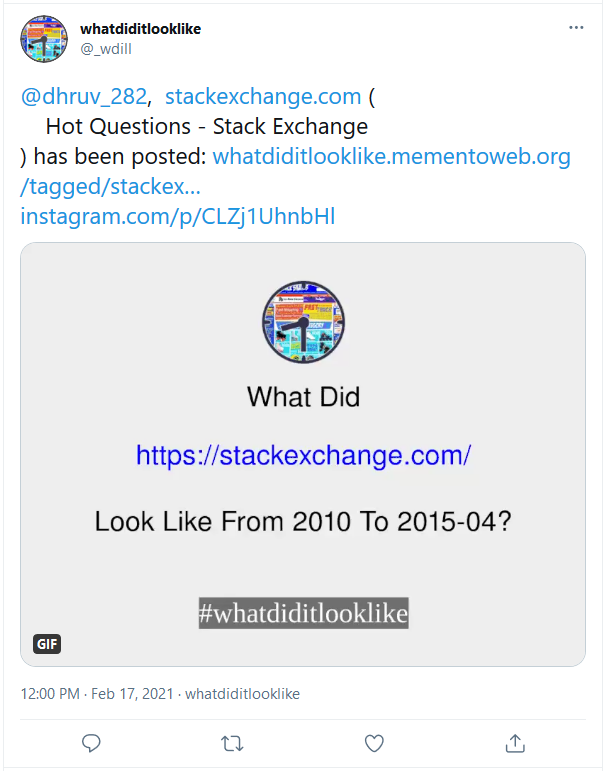}
\caption{Response to the request tweet with timelapse GIF}{\url{https://twitter.com/_wdill/status/1362084422057553923}}
\label{fig:reply_tweet}
\end{figure}

\begin{figure}[ht]
\centering
\includegraphics[width=0.45\textwidth]{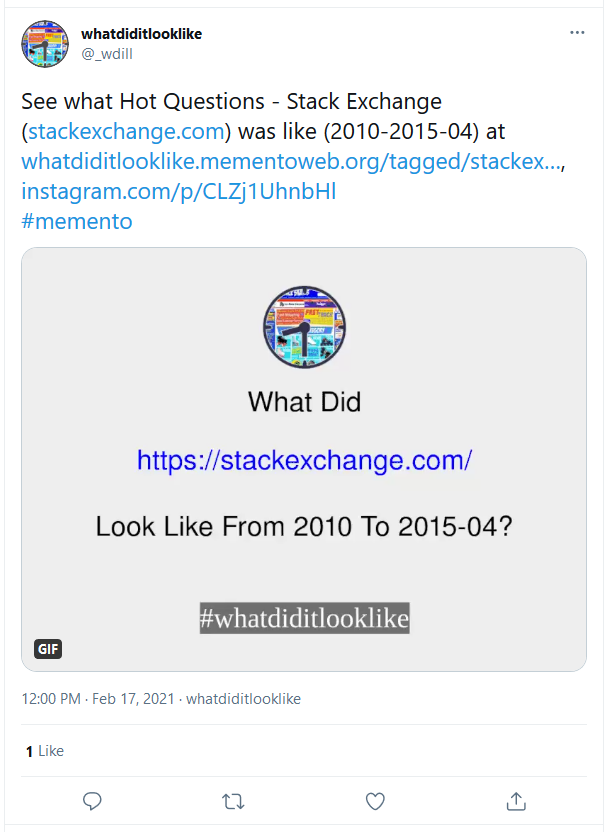}
\caption{Post on timeline for followers with timelapse GIF}{\url{https://twitter.com/_wdill/status/1362084416118382592}}
\label{fig:post_tweet}
\end{figure}

\subsection{Updated Implementation}

The implementation was updated to Python 3 and uses Puppeteer \cite{puppeteer} to capture screenshots. Node.js and Google's Chrome browser installations are required to properly execute the screenshot capture module. \texttt{takeScreenshot.js} is the JavaScript file that handles screenshot captures. 

The first strategy for posting to Instagram was to use the platform's website and this program is located in \texttt{instagram.js}. This program is able to successfully post images to Instagram however video posts are currently unsupported. The \texttt{instagramWithBrowserStack.py} file creates a video post using the Android Instagram app. This file is called by \texttt{common.py} which handles the task of posting the generated output to social media.

\begin{figure*}[ht]
\centering
\includegraphics[width=0.75\textwidth]{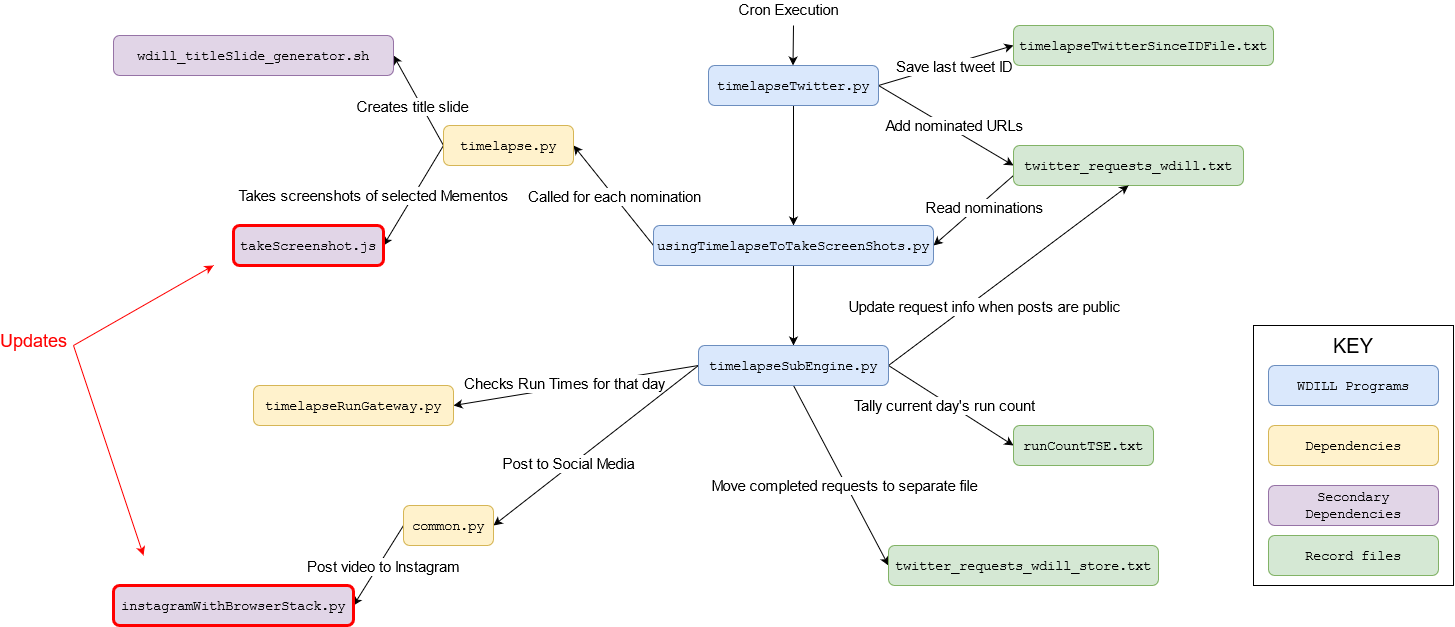}
\caption{Current WDILL program structure}
\label{fig:program-execution-structure}
\end{figure*}

The time-based job scheduler Unix cron is used to set the execution times of the service with the default setting of every hour \cite{cron}. This service is containerized using Docker for easy deployment and avoids any possible conflicts locally \cite{docker}. The base Docker image used for this is \texttt{nikolaik/python-nodejs} since it comes with the latest versions of Python and Node.js \cite{docker-base-img}. Alternatively, the option of running the service is also available for Unix based systems that support the cron utility. The \texttt{setup\_wdill.sh} script handles setting up the necessary files and job execution schedule. Both methods are dependent on proper API credentials and paths set in the \texttt{config} file. This service is currently deployed on a WS-DL provisioned virtual machine in a Docker container on Old Dominion University's Computer Science servers. The updated code base of this project is available on GitHub at \texttt{\url{https://github.com/oduwsdl/wdill}}.

\section{Future Enhancements}

There are certain features and improvements that need to be implemented to make the service more efficient and maintainable.

\subsection{Post to TikTok}

The generated visualization can be shared to the currently popular social media platform, TikTok\footnote{\url{https://www.tiktok.com}}. Implementing this feature was explored by using the TikTok API \cite{tiktok-api}, but it was found that these APIs do not provide functionality to upload a video. Another approach explored was using the Appium framework however solving a captcha is required during the login process which makes it difficult to automate the process. 

\begin{figure}[ht]
\centering
\includegraphics[width=0.45\textwidth]{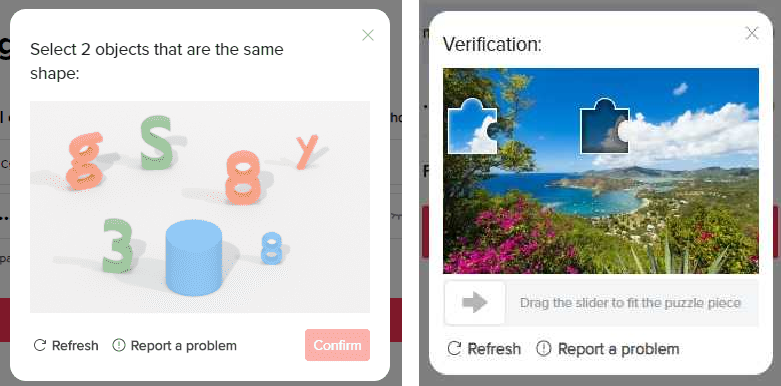}
\caption{Captcha used by TikTok on website(left) and mobile app(right)}
\label{fig:tiktok-captcha}
\end{figure}

\subsection{Front-end system}

Using the past request information that is preserved, charts can be generated for analytics about the timelapse requests. The implemented URL categorization feature can be incorporated in the charts to gain more insight on types of URLs requested. Charts that show the amount of usage of the service over a period of time can also be generated. The data stored in local files would be used to generate these charts which would be accommodated by a front-end system. A maintenance section could also be added to the front-end system which would display the logs generated by the cron job and real-time data for run counts and currently processing requests. This section would also allow for adjusting the service execution schedule and other parameters such as the nomination differential. In order to limit access to this section, appropriate credentials would be needed.

\subsection{Tumblr Blog Search}

The Tumblr Blog the service currently posts to displays posts with latest posts at the beginning. A page that displays all posts in a gallery view with a search functionality could be added to the blog. This would allow users to easily find timelapses before submitting a request via Twitter. This was a previously existing functionality but is currently inactive\footnote{\url{https://web.archive.org/web/20180113093515/http://whatdiditlooklike.mementoweb.org:80/archives}}.

\section{Conclusion}

The objective of this project is to provide a quick and easy way to summarize web page transitions by generating a timelapse. What Did It Look Like is a service that allows users to submit timelapse requests for websites. The Memento framework is used to poll various web archives to access past versions of a requested website. Timelapses are generated using selected mementos and shared to multiple social media platforms. This program is implemented as a single application that relies on multiple external services such as Twitter, Time Travel, and BrowserStack. All interaction with the user is done on Twitter using the provided APIs. What Did It Look Like was originally created in 2015 and has now been updated to use the most current technologies. Enhancements such as date range requests, diversified Memento selection, and sharing to Instagram have been added to the service. The execution is now handled by a time-based job scheduler that allows for execution times to be easily adjusted. The option to run the service in an isolated environment using Docker has been added. There are still some improvements that can be made to the service. For example adding a front-end system to get insight on the type of websites requested and allow adjustments to service parameters.

\bibliographystyle{ACM-Reference-Format}
\bibliography{refs}

\end{document}